\documentclass[prb, twocolumn, nofootinbib]{revtex4-2}
\usepackage{bm}
\usepackage{graphicx}
\usepackage[x11names]{xcolor} 
\usepackage{amsmath}
\usepackage{bbold}
\usepackage{mathrsfs}
\usepackage{color}
\usepackage[unicode=true,colorlinks=true]{hyperref}
\usepackage[capitalize]{cleveref}

\begin{document}

\title{A poor man's theory of circular dichroism in single-wall carbon nanotubes}
\author{S.V.~Goupalov} 
\email{serguei.goupalov@jsums.edu}
\affiliation{Department of Physics, Jackson State University, Jackson MS 39217, USA}

\begin{abstract}
The theory of circular dichroism in single-wall carbon nanotubes derived within the tight-binding method by a complicated approach in previous work is rederived 
in a straightforward way using the multipolar expansion
of the light-matter interaction.
\end{abstract}

\maketitle

\section{Introduction} 
In recent years a significant progress has been achieved in both synthesis and post-growth sorting of chirality-specific single-wall carbon nanotubes~\cite{chemrev}. In particular,
left- and right-handed enantiomers of chiral nanotubes can be distinguished and isolated~\cite{dukovich,peng,ao}. The primary characterization method for analyzing excitonic band structure of
chirality pure nanotubes and determining
enantiomer enrichment and handedness of nanotube populations is the
circular dichroism absorbance spectroscopy~\cite{dukovich,peng,ao,natcom,reich}. 

Sato {\it et al.} have constructed a theory of circular dichroism of single-wall carbon nanotubes within the tight-binding method neglecting excitonic effects~\cite{sato}.
They chose not to employ the multipolar expansion of the light-matter interaction, and the resulting theory turns out to be extremely complex and involving cumbersome derivations. 
Here we show that their results are derivable
directly and easily using the multipolar expansion. 

Within the multipolar expansion, circular dichroism arises as an effect of interference originating from the superposition of transition amplitudes for electric-dipole and magnetic-dipole couplings~\cite{mqed}.
While for chemists single-wall carbon nanotubes represent giant molecules, for physicists they are tiny crystals with one-dimensional periodicity. It has been demonstrated~\cite{svg2005,gzp} that
a direct application of basic notions of solid state physics allows one to obtain, within the tight-binding method, elegant analytical expressions for the electric dipole moment of optical transitions in 
single-walled carbon nanotubes valid for arbitrary nanotube chirality and polarization of light. This enables one to come up with similar expressions for the magnetic dipole moment as well, which makes our task
fairly easy. 

For simplicity, we will restrict our consideration by semiconductor nanotubes.

\section{Zone folding for carbon nanotubes}

Within the tight-binding method, the Bloch wave function
describing an electron state with the band index $s$ and quasi-wave vector
${\bf k}$ in graphene can be written as~\cite{svg2005,sdd}
\[
\Psi_{s,{\bf k}}({\bf r})=\frac{1}{\sqrt{N_{uc}}} \sum\limits_b
C_b(s,{\bf k}) \, \sum\limits_{{\bf R}_b}
e^{i {\bf kR}_b} \, \Phi_{b,p_z}({\bf r}-{\bf R}_b) \,,
\]
where $N_{uc}$ is the number of unit cells (half the number of atoms), $b=A,B$;
${\bf R}_b$ denote the equilibrium atom positions and
$\Phi_{b,p_z}({\bf r})$ are the
$p_z$-atomic orbitals forming the $\pi$-atomic bond.

A graphene sheet can be rolled up to yield a carbon nanotube with the chiral indices $(n,m)$~\cite{sdd}. The electron's
two-dimensional quasi-wave vector ${\bf k}$
has the components $K_1$ along the nanotube's circumference and $K_2$ along
the nanotube's cylindrical axis. In a nanotube $K_1$
becomes quantized: $K_1^{\mu}=\mu/R$, where $R=a \, \sqrt{n^2+m^2+n \, m}/2 \pi$ is the nanotube's radius, $a$ is the lattice constant of graphene. The index $\mu$ can take integral values from $0$ to ${\cal N}-1$,
where ${\cal N}=2(n^2+m^2+n \, m)/d_R$ is the
number of hexagons within the nanotube unit cell, $d_R$ is the greatest common
divisor of $2 n+m$ and $2 m+n$.
The values of $K_2$ are usually restricted by the first Brillouin zone of the nanotube limited by $\pm \pi/T$, where $T=2 \pi \sqrt{3} R/d_R$ is the period of the quasi one-dimensional crystal.

For a semiconductor 
nanotube, the $K$ and $K'$ points in the reciprocal space of graphene with the $(K_1,K_2)$ coordinates
of $({\cal N}/3R,0)$ and $(2 {\cal N}/3R,0)$ respectively 
determine the lowest subbands in the conduction band (as well as the highest subbands in the valence band).
These subbands have $\mu$ indices
equal to the nearest integer numbers to ${\cal N}/3R$ and $2 {\cal N}/3R$ and originate from the $K$ and $K'$ valleys of graphene, respectively.
Optical transitions between the highest subbands in the valence band and the lowest subbands in the conduction band are usually denoted as the $1 \rightarrow 1$ transitions with the optical gap at $E_{11}$~\cite{sato}.

Within the zone-folding scheme the coefficients 
\[
\hat{C}(s,{\bf k}) \equiv {C_A(s,{\bf k}) \choose C_B(s,{\bf k})} 
\]
with $s=c,v$ for the conduction and valence bands respectively are given by~\cite{gzp,sdd}
\[
\hat{C}(c,{\bf k})=\frac{1}{\sqrt{2}} \,
{e^{i \varphi({\bf k})} \choose
1} \,, \,
\hat{C}(v,{\bf k})=\frac{1}{\sqrt{2}} \, {-e^{i \varphi({\bf k})} \choose
1} \,,
\]
where
\[
\varphi({\bf k}) \equiv \varphi(\mu, K_2)=\left\{
\begin{matrix}
\arctan{\frac{\cal B}{\cal A}}, & {\cal A}>0
\cr
\mbox{}
\cr
\arctan{\frac{\cal B}{\cal A}}+\pi, & {\cal A}<0
\end{matrix} \right. \,,
\]
\[
{\cal A}= 2 \, \cos{(k_x a/2 \sqrt{3})} \,
\cos{(k_ya/2)} + \cos{(k_x a /\sqrt{3})} \,,
\]
\[
{\cal B}=2 \, \sin{(k_x a/2 \sqrt{3})} \,
\cos{(k_ya/2)} - \sin{(k_x a /\sqrt{3})}
\,,
\]
$k_x=K_1^{\mu} \, \cos{\alpha} - K_2 \, \sin{\alpha}$,
$k_y=K_1^{\mu} \, \sin{\alpha} + K_2 \, \cos{\alpha}$,
and the angle $\alpha$ is related
to the nanotube chiral angle $\theta$ by $\alpha=\pi/6-\theta$. For a nanotube with the chiral indices $(n,m)$
\[
\cos{\theta}=\frac{2n+m}{2 \, \sqrt{n^2+m^2+n \, m}} \,.
\]
The energy dispersion in the conduction band is determined by
\[
E_c(\mu,K_2)=\gamma_0 \sqrt{1+4 \cos{\frac{\sqrt{3} k_x a}{2}} \cos{\frac{k_y a}{2}} +4 \cos^2{\frac{k_ya}{2}}}  \,,
\]
where $\gamma_0=3$~eV is the transfer integral of the tight-binding method, $E_v(\mu,K_2)=-E_c(\mu,K_2)$.

\section{Circular dichroism spectrum}

Circular dichroism is determined by the difference in absorption rates for left- and right-circular polarized light, averaged over orientations of nanotubes.
It can be expressed as~\cite{mqed}: 
\begin{widetext}
\begin{equation}
\label{cd1}
\langle \Gamma^{(L)} \rangle - \langle \Gamma^{(R)} \rangle =-i \, \frac{16 \pi^2 \, N \, n_{ph} \, \omega \, e}{3 V} \sum\limits_{\mu,\nu,K_2}
\langle v, \mu, K_2 | {\bf r} | c, \nu,K_2 \rangle \, \langle c, \nu, K_2 | {\bf M} | v, \mu, K_2 \rangle \, \delta[\hbar \omega -E_c(\nu,K_2)+E_v(\mu,K_2)] \,,
\end{equation}
\end{widetext}
where $\omega$ is the frequency of incident light, $n_{ph}$ is the number of photons in the incident beam, $N$ is the number of nanotubes, $V$ is the volume,
$e<0$ is the electron charge, ${\bf r}$  and ${\bf M}$ are respectively the operators of the coordinate and the magnetic dipole moment.

\section{Magnetic dipole matrix elements}
Since spin-orbit interaction in carbon nanotubes is negligible, the operator of magnetic moment is determined by the electron orbital motion:
\[
{\bf M}=\frac{e}{2m_ec} \, [ {\bf r} \times {\bf p}] \,,
\]
where $c$ is the speed of light in vacuum, ${\bf p}$ is the momentum operator, and $m_e$ is the free electron mass.
Using the expressions for the coordinate and velocity matrix elements from Ref.~\cite{gzp} one obtains 
\begin{widetext}
\begin{equation}
\langle c, \nu, K_2|M_z|v, \mu, K_2 \rangle=i \, \delta_{\nu,\mu} \, \frac{e \, R^2}{2 \, c \, \hbar} \left\{ E_c(\mu+1,K_2) \, \sin{ \left[\varphi(\mu+1,K_2)-\varphi(\mu,K_2) \right]} \right.
\end{equation}
\[
\left.
-E_c(\mu-1,K_2) \, \sin{[\varphi(\mu-1,K_2)-\varphi(\mu,K_2)]} \right\} \,,
\]
\begin{equation}
\langle c, \nu, K_2|M_x \pm i \, M_y|v, \mu, K_2 \rangle=\mp \delta_{\nu,\mu \pm 1} \, \frac{e \, R}{2 \, c \, \hbar} \, e^{i \, [\varphi(\mu,K_2)-\varphi(\nu,K_2)]/2} \, \cos{\frac{\varphi(\mu,K_2)-\varphi(\nu,K_2)}{2}}
\end{equation}
\[
\times
\left\{ \frac{\partial \varphi(\nu,K_2)}{\partial K_2} \left[ E_c(\mu,K_2)-E_c(\nu,K_2) \right] - 2 \, \frac{\partial \varphi(\mu,K_2)}{\partial K_2} \, E_c(\mu,K_2) \right\} \,,
\]
where the $z$ axis is along the nanotube.
An explicit expression for $\frac{\partial \varphi(\mu,K_2)}{\partial K_2}$ may be found in Ref.~\cite{gzp}.

\begin{figure*}[h]
  \centering
    \includegraphics[width=\textwidth]{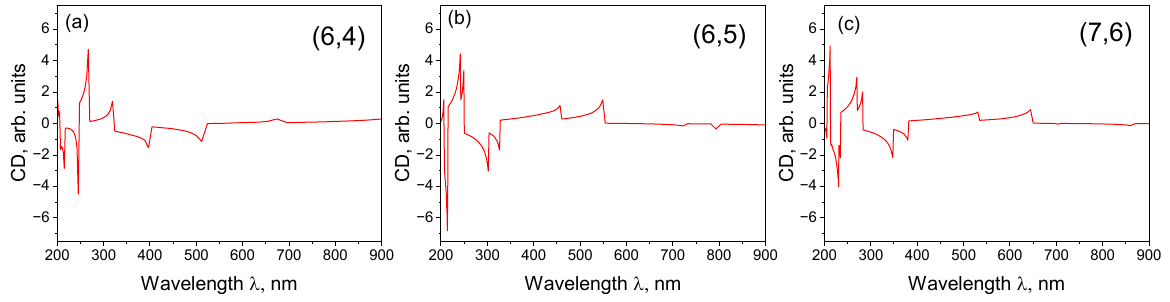}
\caption{Calculated circular dichroism spectra for (a) (6,4) nanotube, (b) (6,5) nanotube, and (c) (7,6) nanotube.}
\label{fig1}
\end{figure*}

\section{Optical rotatory strength}
The optical rotatory strength is a pseudoscalar taking opposite signs for enantiomers~\cite{mqed}. In our case it is given by
\begin{equation}
\label{ors}
-i \, \langle v, \mu, K_2 | {\bf r} | c, \nu,K_2 \rangle \, \langle c, \nu, K_2 | {\bf M} | v, \mu, K_2 \rangle \\
\end{equation}
\begin{align}
&&=
 - \frac{e \, R^2}{8 \, c \, \hbar} \, \delta_{\mu,\nu-1} \, 
\sin{ \left[\varphi(\mu,K_2)-\varphi(\mu+1,K_2) \right]} \left\{ \frac{\partial \varphi(\mu+1,K_2)}{\partial K_2} \left[ E_c(\mu,K_2)-E_c(\mu+1,K_2) \right] - 2 \, \frac{\partial \varphi(\mu,K_2)}{\partial K_2} \, E_c(\mu,K_2) \right\} \nonumber\\
&& + \frac{e \, R^2}{8 \, c \, \hbar} \, \delta_{\mu,\nu+1} \, 
\sin{ \left[\varphi(\mu,K_2)-\varphi(\mu-1,K_2) \right]} \,
\left\{ \frac{\partial \varphi(\mu-1,K_2)}{\partial K_2} \left[ E_c(\mu,K_2)-E_c(\mu-1,K_2) \right] - 2 \, \frac{\partial \varphi(\mu,K_2)}{\partial K_2} \, E_c(\mu,K_2) \right\} \nonumber\\
&& + \frac{e \, R^2}{4 \, c \, \hbar} \, \delta_{\mu,\nu} \, \frac{\partial \varphi(\mu,K_2)}{\partial K_2}
\,
\left\{ E_c(\mu+1,K_2) \, \sin{ \left[\varphi(\mu+1,K_2)-\varphi(\mu,K_2) \right]} 
-E_c(\mu-1,K_2) \, \sin{[\varphi(\mu-1,K_2)-\varphi(\mu,K_2)]} \right\} \,. \nonumber
\end{align}
\end{widetext}

\section{Numerical results}
Substituting the optical rotatory strength, Eq.~(\ref{ors}), into Eq.~(\ref{cd1}) one gets the circular dichroism spectrum, {\it i.e.} the difference in absorption rates
for left- and right-circular polarized light averaged over nanotube orientations. The $\delta$-function in Eq.~(\ref{cd1}) is to be replaced according to
\begin{equation}
\delta[\hbar \omega -E_c(\nu,K_2)+E_v(\mu,K_2)]=\frac{\delta(K_2-K_2^0)}{\left| \frac{\partial E_c(\nu,K_2^0)}{\partial K_2}-\frac{\partial E_v(\mu,K_2^0)}{\partial K_2}\right|},
\label{singularity}
\end{equation}
where $K_2^0$ satisfies the equation
\[
\hbar \omega -E_c(\nu,K_2^0)+E_v(\mu,K_2^0)=0 \,.
\]
At fixed $\omega$, $\mu$, and $\nu$, this equation can have one or two roots $K_2^0$ (excluding the trivial case when it has none). 
To smear out the singularity caused by the denominator in Eq.~(\ref{singularity}), in our numerical calculations we replace
\[
\left| \frac{\partial E_c(\nu,K_2^0)}{\partial K_2}-\frac{\partial E_v(\mu,K_2^0)}{\partial K_2}\right|
\]
by
\[
\sqrt{\left( \frac{\partial E_c(\nu,K_2^0)}{\partial K_2}-\frac{\partial E_v(\mu,K_2^0)}{\partial K_2}\right)^2 + \left( \frac{0.05 \, \gamma_0 \, T}{\pi} \right)^2} \,.
\]
The calculated circular dichroism spectra are shown in Fig.~\ref{fig1} as functions of the wavelength of light for (6,4), (6,5), and (7,6) semiconductor single-wall carbon nanotubes. They turn out to be very similar to the spectra calculated in
Ref.~\cite{sato} (see Fig.~5,~(a)~--~(c) of that paper). 

\section{Conclusions}
We have given a straightforward derivation of the circular dichroism spectra of semiconductor single-wall carbon nanotubes within the tight-binding method, neglecting excitonic effects, and provided
an independent verification
of the results by Sato {\it et al.}~\cite{sato}. Our formalism can serve as a starting point when accounting for excitonic effects.

\acknowledgments
This work was initiated during the author's summer visit to Los Alamos National Laboratory. The author is indebted to Jennifer A. Hollingsworth and the Center for Integrated Nanotechnologies for hospitality
and to Andrei Piryatinski for stimulating discussions.
The work was supported in part by the U.S. Department of Energy, Office of Science, Office of Workforce Development for Teachers and Scientists (WDTS) under the Visiting Faculty Program (VFP) and, 
in part, by NSF through DMR-2100248. 

\end{document}